# Power Installations based on Activated Nuclear Reactions of Fission and Synthesis

Yu.V. Grigoriev [1,2], A.V. Novikov-Borodin [1]

[1] Institute for Nuclear Research RAS, Moscow, Russia
[2] Joint Institute for Nuclear Research, Dubna, Russia

**Abstract.** The general scheme of power installations based on nuclear reactions of fission and synthesis activated by external sources is analyzed. The external activation makes possible to support nuclear reactions at temperatures and pressures lower than needed for chain reactions, so simplifies considerably practical realization of power installations. The possibility of operation on subcritical masses allows making installations compact and safe at emergency situations. Installations are suitable for transmutation of radioactive nuclides, what solves the problem of utilization of nuclear waste products. It is proposed and considered schemes of power installations based on nuclear reactions of fission and fusion, activated by external sources, different from ADS systems. Variants of activation of nuclear reactions of fission (U-235, 238, Pu-239) and fusion (Li-6,7, B-10,11) are considered.

## Introduction

World power requirement constantly increases and apparently there is no alternative to the nuclear power engineering, but existing power stations based on fission reactions meet serious ecological, economical and safety problems, Ref. [1-4]. Nowadays, there are hundreds of tons of long-living fission fragments and minor actinides which need to be kept in special radioactive waste storages or to be transmuted into short-living isotopes and the acuteness of this problem is permanently increasing.

Some of these problems may be solved with help of the accelerator-driven systems (ADS), Ref. [5,6], where nuclear reactions are activated with help of charged particle accelerators. High intensity beams of protons with energies from 500 MeV to 1.5 GeV knock out neutrons from heavy metal target in a result of hadron-nuclear cascade. The neutron flux activates nuclear reactions in subcritical blanket with fissile materials inside the reactor. It may be used as for 'after-burning' of long-lived radioactive nuclei to utilize radioactive waste products, Ref. [5,6], as for activation of nuclear reactions inside powerful reactors (ADSR), Ref. [7-9]. Powerful accelerators used in ADSRs predefine their processing and operating complexity, also as low efficiency in comparison with existing atomic power stations. Moreover, ADS may solve only a small part of existing problems.

Thermonuclear stations based on fusion reactions do not need radioactive fuel and would solve many problems of nuclear power engineering, but extreme conditions needed for chain deuterium-tritium fusion reactions, Ref. [10-12], are achieved now only during thermonuclear bomb explosion, while the problem of maintaining them long enough in thermonuclear reactors is not solved yet. Thermonuclear stations are not realized and are under development.

Extra-high temperatures $\sim 10^7$-$10^8$ K needed for chain fusion reactions correspond to only 1-2 MeV energy of reaction components. It looks much easier to accelerate one of the reaction components to 1-2 MeV to activate fusion reactions instead of heating all fuel in the reactor. Nuclear fusion reactions would be maintained and controlled by permanent injection of activating particles, but one needs to find such fusion reactions which could be effectively activated.

General operational efficiency of installations with nuclear reaction activation is analyzed in this paper. Nuclear fusion reactions available for effective activation are considered and some methods of such activation are analyzed. There are proposed the variants of power installations based on nuclear fusion reactions activated by sources, which uses the fission reactions. Such nuclear-thermonuclear reactors and sources, by means of decreasing an amount of used nuclear fission fuel and, therefore, an amount of radioactive waste products, can solve many ecological problems of nuclear power engineering and also can increase its efficiency and safety.

## 1. General scheme of NRA installations

The principal scheme of installations with nuclear reaction activation (INRA) is presented on Figure 1A. The activator initiates and maintains in the reactor the nuclear reactions going with energy output, the extractor transports produced energy to converter, which transforms it to convenient form. Some part of INRA power is used by consumers and another one for INRA operation. The INRA power may be controlled by varying the parameters of activation, necessary regimes of operation may be maintained by the feedback.

Design of NRA installation presented on Figure 1A minimizes total losses in it, because the activator is placed inside the reactor, which is inside the extractor, so nor heat energy nor energy of particles escaped from activator or reactor are not lost, but is accumulated in the extractor, which simultaneously shields from radiation. Exactly such design is used in modern projects of ADS reactors, such as MYRRHA in Belgium and CLEAR-I in China, Ref. [8,9].

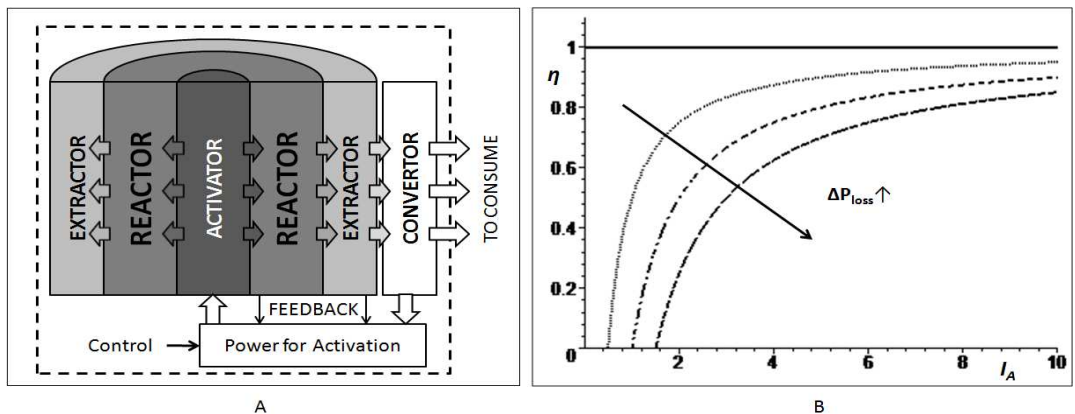

**Figure 1** General scheme of NRA installations (A) and their efficiency versus activation (B)

The NRA efficiency ($\eta$) may be defined as a relation of useful ($P_U$) (to consumer) and total ($P$) power of the reactor: $\eta=P_U/P$. Total power losses will be defined as: $\Delta P=P-P_U$, so efficiency will be expressed as:

$$\eta = P_U / P = 1 - \Delta P / P = 1 - (\Delta P_1 + \Delta P_2 + ... + \Delta P_n)/P, \qquad (1)$$

where $\Delta P_i$ are losses in INRA parts: in activator, reactor, extractor, convertor, etc.

The total power ($P$) of the reactor consists of kinetic energy of activating particles ($\varepsilon_A$) with intensity $I_A$ and an energy output (Q-value) ($\varepsilon_N$) of nuclear reactions (or chains of reactions) activated by some part of activating particles: $I_N=k_L k_N I_A$: $P=k_L \varepsilon_A I_A + \varepsilon_N I_N$. Here $k_L$ is a coefficient characterizing the losses of activating particles during delivering them into the active zone of the reactor, $k_N=k_N(n_N,\sigma_N)$ is a coefficient of nuclear interactions depending on concentration $n_N$ of nuclei of activated reactions and their cross-sections ($\sigma_N$). Thus, total power may be expressed as:

$$P = K_A I_A = q K_A J_A, \quad K_A = k_L(\varepsilon_A + k_N \varepsilon_N), \qquad (2)$$

where $J_A=qI_A$ is a current of activating particles with an electric charge ($q$).

At zero useful power ($P_U=0$) or at zero efficiency ($\eta=0$) it is needed the threshold power: $P_0=\Delta P_0=\Delta P(\eta=0)$, hence, threshold intensity or current: $P_0=K_A I_0=qK_A J_0$ to compensate the total losses in NRA installation in a free running regime. If total losses slightly depend on regimes of INRA operation (as in case of ADS reactors, where accelerator is responsible for a large part of total losses), then $\Delta P \approx P_0$ and the INRA efficiency may be expressed as:

$$\eta = 1 - \Delta P / P \approx 1 - P_0 / P = 1 - I_0 / I_A = 1 - J_0 / J_A. \qquad (3)$$

The dependence of NRA efficiency on the activation ($I_A$, $J_A$) with total losses increasing is shown on Figure 1B. Positive values of threshold power (intensity, current) need to be guaranteed in all regimes of safety operation of INRA, because negative values correspond to chain nuclear reactions in the reactor. During burning-out of the nuclear fuel the concentration of activated atoms ($n_N$) is decreasing. It leads to decreasing the coefficients $k_N$ and $K_A$ and increasing the threshold intensity ($I_0$), so to decreasing the INRA efficiency.

One needs to increase the activation current to increase the efficiency of ADS reactors (see Exp. (3) and Fig.1B) and to use powerful proton accelerators. It increases the total losses and a threshold in ADSR, so needs additional increasing of accelerator intensity to overcome a threshold. Estimations show that economically effective power of ADS reactors needs to be not less than 10 MW. The power of ADS reactor in project CLEAR-I in China is 10 MW and the power of ADS reactor in project MYRRHA in Belgium is 65-100 MW, Ref. [8,9].

## 2. Activation of nuclear reactions

Activation with neutrons may be quite effective, because they do not have charge and mainly take part in nuclear interactions without losing much energy for ionization and scattering. Neutron activation may be easily realized on practice and some variants of practical realization of such activation will be considered in Section 3.

Charged particles, vice versa, activate nuclear reactions to a much lesser degree, because they lose energy mainly for ionization and scattering. Small track length of charged particles in substance (for example, it is only 0.6 mm in aluminum for 10 MeV protons) is also a serious technical problem for delivering them into reaction zone, because accelerator facilities usually need to be isolated from the reactor. Nevertheless, some nuclear reactions activated by charged particles have high Q-value, and, taking into account that the energy of activating particles, which do not take part in nuclear reactions, is not involved in total losses (see Section 1), such activation may also be effective. Some perspective reactions of activation with charged particles will also be considered.

### 2.1 Basic reactions of neutron activation

Let's consider the reactions of interaction of neutrons with nuclei of lithium-6: $^6$Li$(n,\alpha)$T and boron-10: $^{10}$B$(n,\alpha)^7$Li, where T=$^3_1$H is tritium, and $\alpha=^4_2$He is alpha-particle, which go with energy outputs (Q-values) of 4.78 and 2.79 MeV correspondingly:

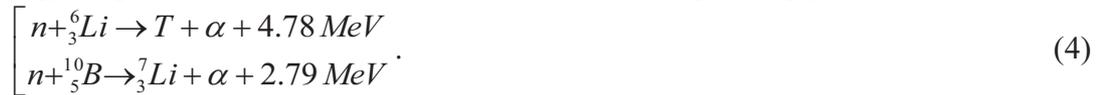

$$\begin{bmatrix} n+^6_3Li \rightarrow T + \alpha + 4.78\,MeV \\ n+^{10}_5B \rightarrow ^7_3Li + \alpha + 2.79\,MeV \end{bmatrix}. \qquad (4)$$

Cross-sections of interactions of neutrons with nuclei of lithium-6 from ENDF database (Evaluated Nuclear Data File, USA) are shown on Figure 2 (the situation with boron-10 is similar). In the energy range from 0 to $10^{-4}$ MeV the total cross-section (curve 1) practically coincides with one of reaction $^6$Li$(n,\alpha)$T from Exp.(4) (curve 3), so this reaction is dominant. In a range from $10^{-4}$ to ~2-3 MeV the elastic scattering becomes comparable, but with losing their energy, neutrons will also take place in reactions Exp.(4). The same situation is for scattering with excitation (curves 5 and 6) on a range 2-3 MeV. Rival reaction: $^6$Li$(n,\gamma)^7$Li (curve 7) has very small cross-section (4-5 degrees less than dominant reactions) and may be neglected, also as reactions: $^6$Li$(n,2n+\alpha)^1$H (curve 4) and $^6$Li$(n,p)^6$He (curve 8) appeared from 3 MeV and 4 MeV correspondingly with cross-sections in 1-2 degrees less than dominant reactions. Thus, activation of reactions Exp.(4) in neutron energy range 0-10 MeV is effective, but a range 0-3 MeV is preferable for activation of reactions Exp.(4).

The Q-values of reactions Exp.(4) seem small in comparison with fission reaction of uranium-235, which in average is about 180 MeV. However, the Q-value per one nucleon for lithium-6 is: 4.78/6=0.80, while for uranium-235: 180/235=0.77 MeV/nucleon. It means that Q-values of equal masses of lithium-6 and of uranium-235 approximately equal to each other, and Q-value of boron-10 is three times less (0.28 MeV/nucleon).

From other side, one neutron is needed to initiate the decay of one nucleus of uranium-235 to get 180 MeV, but it needs 180/4.78=38 neutrons to get the same energy from lithium-6 and 180/2.79=65 neutrons to get it from boron-10, so lithium-6 needs 38 times more intensive neutron flux than uranium-235 to get the same power and boron-10 65 times more. Moreover, the coefficient of neutron multiplication for uranium-235 is 2.4, it is 2.5 for uranium-233 and 238, and 2.9 for plutonium-239, so fission reactions may be self-maintained, while lithium-6 and boron-10 always need quite intensive neutron source for activation.

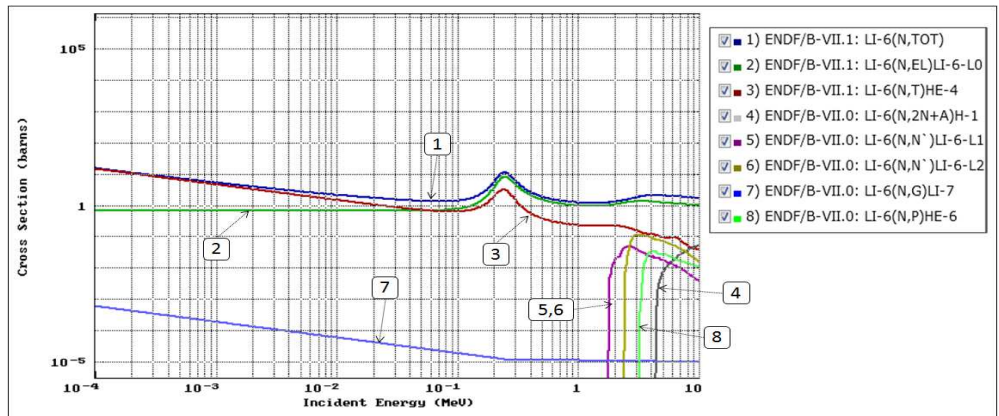

**Figure 2** Cross-sections of interactions of neutrons with Lithium-6

One can improve the situation by adding nuclei of beryllium-9 in lithium-6 to activate neutron reproduction in break up reactions: $^9\text{Be}(\alpha,n+\alpha)^8\text{Be}$ and reactions: $^9\text{Be}(\alpha,n)^{12}\text{C}$:

$$\begin{bmatrix} \alpha + ^9_4Be \rightarrow n + \alpha + ^8_4Be \\ \alpha + ^9_4Be \rightarrow n + ^{12}_6C + 4.44\, MeV \end{bmatrix}. \quad (5)$$

The cross-sections of alpha-particle interactions with beryllium-9 and lithium-6 from EXFOR library (Experimental Nuclear Reaction Data, IAEA) are shown on Figure 3. Multiple neutron production in series of break up reactions is possible, and cross-sections of total neutron production (points 3) of reactions Exp.(5) are very high: 240 barns at 2 MeV and 547 at 4. When neutrons reactivate reactions Exp.(4), a chain of reactions Exp.(4)-(5) is activated. These chain reactions will be considered as basic ones for NBA installations.

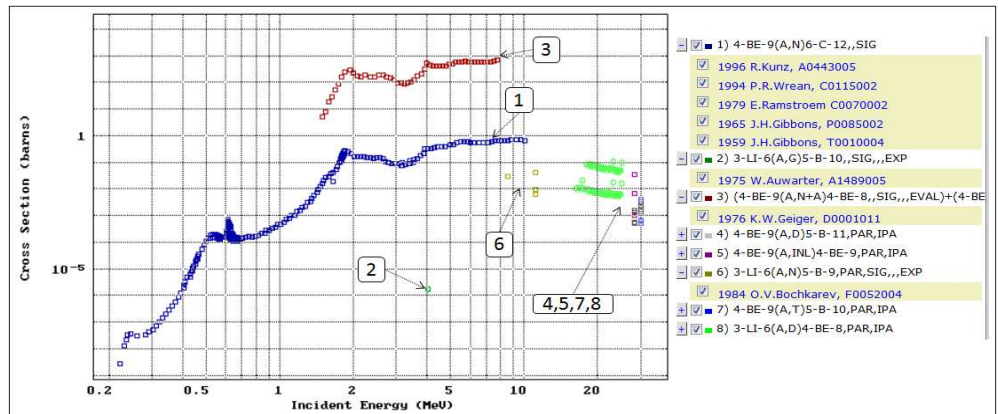

**Figure 3** Cross-sections of alpha-particle interactions with Berillium-9 and Lithium-6

The threshold energy for $(\alpha,n)$ interactions of lithium-6 with alpha-particles is 6.32 MeV (see Table 1, Ref. [13]), and in EXFOR library there are only four experimental points (points 6) around 10 MeV with $10^{-3}$-$10^{-2}$ barns for reaction $^6\text{Li}(\alpha,n)^9\text{B}$, so reactions Exp.(5) are dominant. However, basic reactions are interconnected, so direct experiments of, for example, neutron activation of basic reactions in samples with different concentrations of lithium-6 and beryllium-9 or lithium-6 and boron-10 are extremely needed to estimate the value of multiplication coefficient.

**Table 1** Q-values, threshold energies and Coulomb barriers for ($\alpha,n$) reactions

| Nucleus | Natural Abundance (%) | Q-Value (MeV) | Threshold Energy (MeV) | Coulomb barrier (MeV) | Maximum Neutron Energy for 5.2 MeV Alphas (MeV) |
|---|---|---|---|---|---|
| Li-6 | 7.5 | -3.70 | 6.32 | 2.1 | |
| Li-7 | 92.5 | -2.79 | 4.38 | 2.1 | 1.7 |
| Be-9 | 100 | +5.70 | 0 | 2.6 | 10.8 |
| B-10 | 19.8 | +1.06 | 0 | 3.2 | 5.9 |
| B-11 | 80.2 | +0.16 | 0 | 3.2 | 5.0 |

On practice one needs to take into account that efficiency of activation of basic reactions also depends on geometry and construction of the reactor, because of losses of activating particles, which may escape from the reaction zone. Tritium produced in reactions Exp. (4) with lithium-6 may be useful for thermonuclear energetic, but special technique has to be provided for its removing off the reaction zone.

By using basic reactions Exp. (4)-(5) one can solve many serious problems of existing nuclear power engineering, but additional investigations, especially direct experiments, are extremely needed to determine parameters of these reactions.

### 2.2 Activation by charged particles

Chains of basic reactions Exp. (4)-(5) can also be activated by charged particles, and if the problems of delivering them to the reactor active zone and of small values of their nuclear interaction cross-sections are overcome, they would be quite effective for activation. Here we only point out to some of such capabilities without detailed analysis.

The EXFOR data analysis shows that for interaction of 0.3-5.0 MeV proton with nucleus of lithium-6 the dominant nuclear reaction is (see Figure 4A):

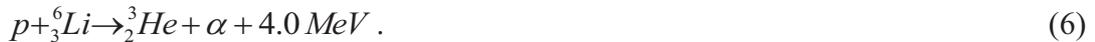

$$p + {}_{3}^{6}Li \rightarrow {}_{2}^{3}He + \alpha + 4.0 \; MeV \; . \tag{6}$$

In energy range 1-2 MeV the cross-section of this reaction is 0.2-0.3 barns. The 4.0 MeV energy output is distributed between a nucleus of helium-3 (~2.3 MeV) and an alpha-particle (~1.7 МэВ), which, if beryllium-9 is added, can activate basic reactions Exp. (4)-(5). In its turn, the 1-3 MeV helium-3 nucleus interact with lithium-6 in two dominant reactions (see Figure 4B): ${}^6Li({}^3He,D){}^7Be$ (~0.4 barns) and ${}^6Li({}^3He,\alpha+p){}^4He$ (~0.04 barns).

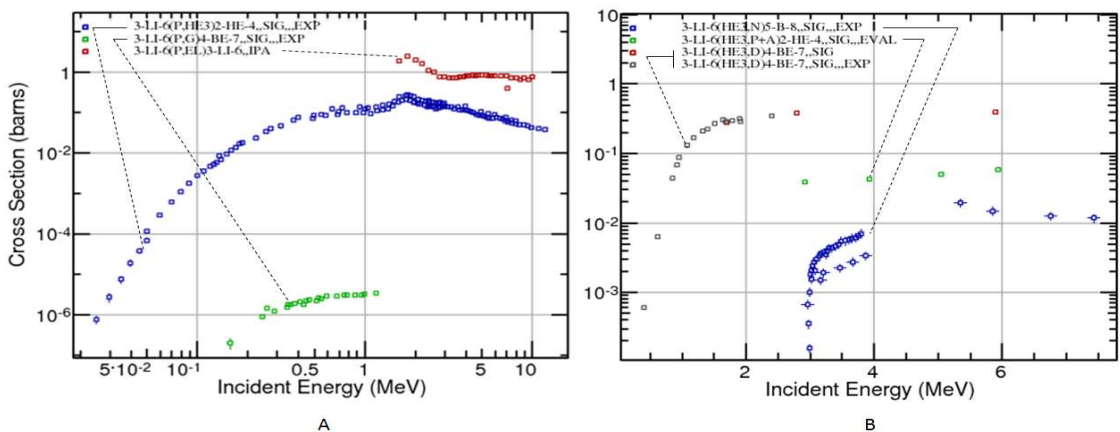

**Figure 4** Cross-sections of interactions of proton (A) and ${}^3He$ (B) with ${}^6Li$

Two alpha-particles produced in reaction ${}^6Li({}^3He,\alpha+p){}^4He$ can activate two chains of basic reactions at ones and proton can maintain the activation through reaction Exp. (6). The deuteron produced in reaction ${}^6Li({}^3He,D){}^7Be$ with large Q-value of 16.9 MeV may also interact with lithium-6 with large 22.4 MeV Q-value and with production of two alpha-particles:

$$D + {}^6_3Li \rightarrow 2\alpha + 22.4\, MeV, \qquad (7)$$

which may activate two chains of basic reactions Exp. (4)-(5). Rival reactions: $^6Li(D,n)^7Be$ and $^6Li(D,p)^7Li$ have Q-values 3.4 and 5.0 MeV correspondingly, and both neutron (through reaction Exp. (4)) and proton (through reaction Exp. (6)) produced in them may also activate chains of basic reactions.

Thus, protons in reaction Exp.(6) and deuterons in reaction Exp.(7) may be very effective for activation of basic reactions Exp. (4)-(5).

In case of boron-10 by adding boron-11 and beryllium-9, one may activate the basic reactions by protons with reaction: $^{11}B(p,2\alpha)^4He$, which has a Q-value of 8.7 MeV and produced alpha-particles may activate three chains of basic reactions at once.

Thus, in spite of small cross-sections of nuclear interactions of charged particles, they have large Q-values and may activate few chains of basic reactions at once, so the activation of basic reactions by protons and deuterons could be quite effective, if mentioned above technical problems of delivering them into reaction zone are solved.

### 3. Practical realization

Basic reactions Exp. (4)-(5) may be activated by different particles: neutrons, alpha-particles, protons and deuterons, so many existing isotopic sources (see Ref. [13,14]) may be used as activators.

Some intensive spontaneous fission neutron sources are presented in Table 2, Ref. [13]. The most intensive is californium-252. It has high neutron multiplication coefficient (3.757) and one gram of it emits about $2.34 \cdot 10^{12}$ neutrons per second, but it is expensive and has short half-life 2.645 years. For other isotopes one can get much better results by using their alpha-activity and $(\alpha,n)$ reaction, usually with beryllium-9.

**Table 2** Spontaneous fission neutron yields

| Isotope | Total half-life | Spontaneous fission yield $(n.s^{-1}.g^{-1})$ | Spontaneous fission multiplicity |
|---|---|---|---|
| Pu-238 | 87.74 y | $2.59 \cdot 10^3$ | 2.21 |
| Pu-240 | 6560 y | $1.02 \cdot 10^3$ | 2.16 |
| Cm-242 | 163 d | $2.10 \cdot 10^7$ | 2.54 |
| Cm-244 | 18.1 y | $1.08 \cdot 10^7$ | 2.72 |
| Bk-249 | 320 d | $1.00 \cdot 10^5$ | 3.40 |
| Cf-252 | 2.646 y | $2.34 \cdot 10^{12}$ | 3.757 |

Some characteristics of $^9Be(\alpha,n)$ sources activated by the alpha-active isotopes are presented in Table 3, Ref. [13]. One can see that in this case the neutron yield from the same isotopes is few orders of magnitude more than in previous case. For example, neutron yield from Pu-238 is $2 \cdot 10^4$ times more, from Cm-242 $10^3$ times, etc.

**Table 3** Characteristics of $^9Be(\alpha,n)$ sources

| Nuclide | Half-life | $E_\alpha$ (MeV) | Yield per $10^6$ alpha's $(n.s^{-1})$ | Yield $(n.s^{-1}.g^{-1})$ | Average neutron energy (MeV) | γ-dose at 1 m for $10^6$ n/s in mGy/h |
|---|---|---|---|---|---|---|
| Pu-238 | 89 y | 5.50 | - | $4.5 \cdot 10^7$ | 4.0 | <0.01 |
| Pu-239 | 24110 y | 5.14 | 65 | $1.2 \cdot 10^5$ | 4.59 | ≤0.01 |
| Po-210 | 138 d | 5.30 | 73 | $1.1 \cdot 10^{10}$ | 4.54 | <0.001 |
| Am-241 | 433.6 y | 5.48 | 82 | $6.5 \cdot 10^8$ | 4.46 | 0.01 |
| Cm-242 | 163 d | 6.10 | 118 | $\sim 1 \cdot 10^{10}$ | 4.16 | <0.01 |
| Cm-244 | 18.1 y | 5.79 | 100 | $2.5 \cdot 10^8$ | 4.31 | <0.01 |
| Ra-226 | 1620 y | 7.69-4.77 | 502 | $1.5 \cdot 10^7$ | 3.94 | 0.5 |
| Ac-227 | 22 y | 7.36-5.65 | 702 | $1.700 \cdot 10^9$ | 3.87 | 0.07 |

One may activate basic reactions Exp.(4)-(5) by adding neutron- or alpha-active isotopes directly into Li-Be or B-Be mixture, but, as it was mentioned before, this solution needs additional investigations and direct experiments.

Probably, not the most efficient, but more simple way is to separate activating sources from a zone of basic reactions. Two variants of design of such autonomous energy sources operating as heat generators are presented on Figure 5A. Here (1) are lithium-6 or boron-10 with beryllium-9, (2) is a neutron moderator, (3) is the activating neutron isotope source, (4,5) are neutron reflector and absorber, (6) is a source vessel. Design with reflector and absorber is intended for autonomous operation, while design without them is for operation in common vessel with reflector, moderator and biological shielding. The last design gives a possibility to control the power of heat-generation by changing the mutual position of rods. When rods are close to each other the intensity of activating neutron increases considerably due to mutual influence of rods and heat-generation is increasing non-linearly. The design with common vessel looks similar to existing reactors of nuclear power stations.

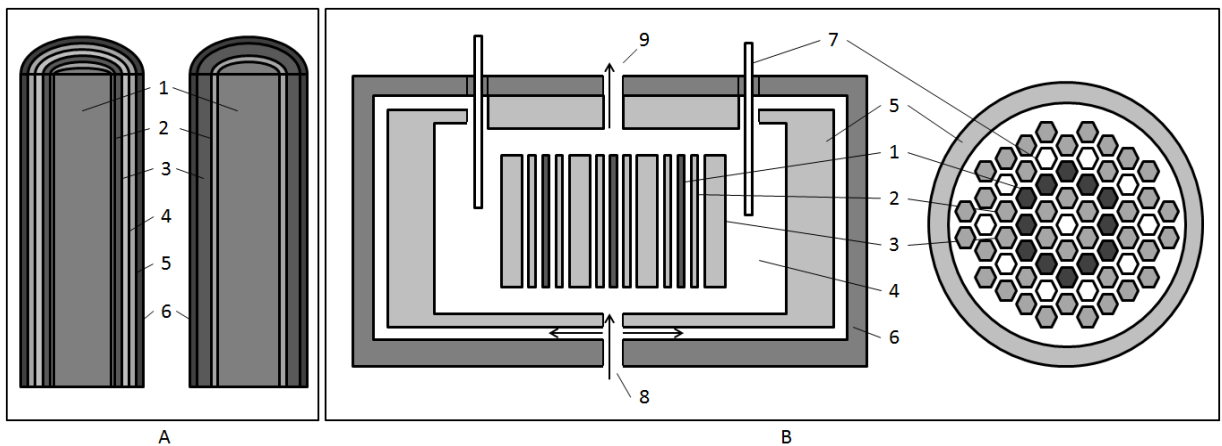

**Figure 5** Designs of autonomous energy sources (A) and heterogeneous nuclear-thermonuclear reactors (B)

Modification of existing nuclear power stations is one of the perspective ways to use nuclear fusion energy of basic reactions Exp. (4)-(5) on practice. Proposed design of such nuclear-thermonuclear reactor (NTR) based on modified heterogeneous atomic reactor and the geometry of its active zone (view from above) are shown on Figure 5B. Here (1) are rods with uranium or plutonium, (2) rods with lithium-6 or boron-10 with beryllium-9 or without it, (3) neutron moderators, (4) the coolant, (5) the neutron reflector, (6) the reactor vessel, (7) the control rods, (8,9) coolant inlet and outlet.

Activating rods (1) with uranium or plutonium, used in existing heterogeneous reactors as fuel elements, in NTR operate as activators of basic reactions in basic fuel rods (2) containing lithium-6 or boron-10 with beryllium-9 or without it. Basic rods with beryllium are used as main fuel elements with neutrons reproduction, while basic rods without beryllium, being also fuel elements, are neutron absorbers. When activating and basic rods have the same sizes, one may easily replace them with each other to modify, if necessary, geometry of active zone of the reactor. Neutrons, produced in reactions Exp. (5), are fast with energies about 1-3 MeV, so both thermal and fast neutron reactors may be modified to be used in nuclear-thermonuclear operational regime.

Probably, it would be more effective to use universal fuel rods, where activators (uranium or plutonium) are placed together with fuel materials (lithium or boron with beryllium), but as it was mentioned before, such variants need additional investigations.

In nuclear-thermonuclear reactors, operated on fission and fusion reactions, one may decrease the amount of radioactive fuel, and in many respects may solve the ecological problems concerning the extraction and mining of the nuclear fuel and burial of radioactive wastes.

## Conclusion

General operational efficiency of installations with activation of nuclear reactions were analyzed. Nuclear fusion reactions with high Q-values and methods of activation of these reactions were considered. There were proposed the variants of practical realization of installations with activation of nuclear fusion reactions: the design of autonomous energy sources (heat generators) activated by intensive isotopic neutron sources and the design of nuclear-thermonuclear reactors modified from existing nuclear power stations. Proposed nuclear-thermonuclear reactors and autonomous heat generators do not need extra-high temperatures and pressures required for maintaining the fusion reactions in thermonuclear reactors, so they may be much easier realized on practice. Radioactive fuel (isotopes of uranium or plutonium) in them is used only for activation. It decreases an amount of radioactive fuel and, therefore, an amount of radioactive waste products; so many serious ecological and safety problems of nuclear power engineering may be solved.

Additional investigations are extremely needed to determine parameters of basic reactions. Especially, one needs direct experiments on activation of basic reactions in lithium-beryllium or boron-beryllium mixture with different concentrations. A whole complex of R&D also needs to be carried out for practical realization of proposed installations.